# Contributions to the Theory of Thermostated Systems II: Least Dissipation of Helmholtz Free Energy in Nano-Biology

Ronald F. Fox
Regents' Professor Emeritus
School of Physics
Georgia Institute of Technology
Atlanta, Georgia

## ABSTRACT

In this paper, we develop further the theory of thermostated systems along the lines of our earlier paper ([1] Fox). Two results are highlighted: 1) in the Markov limit of the contracted description, a *least dissipation of Helmholtz free energy principle* is established; and 2) a detailed account of the appropriateness of this principle for nano-biology, including the evolution of life, is presented.

## Introduction

In ([1] Fox), hereafter referred to as TSI, we referred to the extensive work done over the past 25 years on thermostated systems. In some of that work computer simulations were performed and it was often the case that the trajectory was stopped suddenly and then instantaneously all momenta were reversed and the reversed trajectory was followed and compared with the original trajectory. Rather elegant theoretical results were achieved from this perspective, including, among others, the Jarzynski equality and the Crooks theorem (for references, see TSI). In TSI we noted that these instantaneous momentum reversals of each and every particle are physically impossible to achieve in the laboratory. Instead of following that approach we sought an approach in which all concepts are realizable physically. We stated that from the perspective of nano-biology applications, the major constituent of these systems is water that acts as an *interstitial thermal reservoir* with high heat capacity and strong heat conductivity. A nano-biological system can be looked at instead as comprised of all of its solute molecules with the water molecules producing Brownian motion on the solute molecules but otherwise the water molecules are not explicitly represented. In nano-biology we typically have very low Reynolds number and viscosity overwhelms inertia ([2] Fox, [3] Berg). We call such systems strongly over-damped. In a second step in TSI we contracted the description by integrating out all of the solute momenta. This is an objective procedure, free of subjective coarse graining. The resulting *contracted description* describes the *spatial evolution* of the solute molecules in the form of a non-Markov evolution equation, given by equation (43) of TSI. Two techniques were used to achieve the contraction, projection operators and boson operator representations ([4] Steiger and Fox, [5] Fox). The resulting description captures what we actually see, spatially, and not the unseen underlying momenta dependent Hamiltonian dynamics.

The strongly over-damped case permits a further reduction of the description that was used in TSI to obtain the Markov approximation given by equation (125) of TSI. This Markov version of the contracted description is the focus of the present paper. In section **I** the explicit generalization to $n$ interacting solute particles, each in 3 spatial dimensions, is given. The proof that the *macroscopic* Helmholtz free energy, $F$, is monotonically decreasing is made explicit for this generalization. In section **II** we prove a new result for these systems. We show how to use the conditional probability solution, $P_2$, for the Markov limit of the contracted description to construct *microscopic* paths in solute coordinate space that are reminiscent of the time reversible phase space trajectories used in the earlier research of others. $P_2$ is the Green's function solution to the Markov limit of the contracted description. Using the Trotter Product formula, together with the Dirac delta function Green's function for a simultaneously first order in time and first order in space partial differential equation, we obtain an explicit rendering of the probability of a microscopic path in the strongly over-damped regime. This enables us to establish that the *microscopic Helmholtz free energy dissipation for the most probable path is least compared to all other paths starting from the same initial point.*

Section **III** is dedicated to interpreting these results in the context of nano-biology. We begin by looking at the details for ubiquinone, a fundamental component of electron transport chains. With it we can assess the sizes of $\Delta t$ and $dr$ as well as the conditions for the Markov limit of the contracted description. For example, $\Delta t$ needs to be small enough for validity of the Trotter Product formula and simultaneously large enough for the Markov limit. A number of questions may be asked about the contemporary structure of cells and of their metabolic pathways in light of the *least dissipation principle*. One may well ask: Why is energy metabolism, based on the oxidation of sugar (glucose) by oxygen, broken up into many small steps that harvest the available free energy in small useful packets of energy in the form of ATP? Why not simply burn sugar with a flame and get the decrease of free energy and the approach to full equilibrium over with rapidly? Another issue is: What happens to the ATP? It could simply hydrolyze and turn its energy into heat, reaching full equilibrium quickly. Instead it activates a host of other molecules so that they in turn can react or change conformation. Principal among the activated species are amino acids activated for condensation into proteins instead of simple hydrolysis of the ATP or of the activated monomers, the amino-acyl-adenylates. Similarly for the polymerization of mononucleotides into polynucleotides, such as RNA and DNA, the monomers initially are activated by ATP. Why does DNA get replicated or transcribed using activated monomers instead of these activated nucleotides and the DNA simply hydrolyzing instead? We can also ask why proteins produced in the cell self-assemble into complexes instead of hydrolyzing first. These are a few of the examples that become elucidated through application of the *least dissipation of free energy principle*.

## I. Monotonic decrease of the macroscopic Helmholtz free energy

In TSI the presentation was given in one spatial dimension in order to make the presentation as transparent as possible. We now expand the description to include $n$ interacting particles, each in 3 spatial dimensions. Sections **IX** and **X** of TSI are the relevant sections. The macroscopic Helmholtz free energy, $F$, is defined by
(1)

$$F = E - TS$$

in which $T$ is the thermostat temperature, $E$ is the internal energy and $S$ is the entropy. For the particle labeled by $j$ the coordinates are given by $\boldsymbol{r_j}$. The $n$ particle distribution function at time $t$ is written as $R(\boldsymbol{r_1}, \boldsymbol{r_2}, \dots, \boldsymbol{r_n}, t)$. The potential energy, previously simply denoted by $U$, will now be given by
(2)

$$U = \sum_{j=1}^{n} U_j + \frac{1}{2} \sum_{j,k=1}^{n} U_{jk}$$

in which the $U_j's$ are one particle potentials that include the walls of the container, and the $U_{jk}'s$ are two particle interaction potentials. The factor of ½ corrects for double counting the pair potentials. Inclusion of multi-particle potentials for 3 or more particles is not excluded in principle, and is indeed necessary if one is to model reactions and catalysis purely classically, without help from quantum mechanics. Given these preliminaries, the internal energy in equation (1) is given by
(3)

$$E = \int d\boldsymbol{r_1} d\boldsymbol{r_2} \dots d\boldsymbol{r_n} \left( \sum_{j=1}^{n} U_j + \frac{1}{2} \sum_{j,k=1}^{n} U_{jk} \right) R(\boldsymbol{r_1}, \boldsymbol{r_2}, \dots, \boldsymbol{r_n}, t)$$

in which the differentials are meant to denote differential volumes: $d\boldsymbol{r} = dxdydz$. The entropy, $S$, is given by
(4)

$$S = -k_B \int d\boldsymbol{r_1} d\boldsymbol{r_2} \dots d\boldsymbol{r_n}\, R(\boldsymbol{r_1}, \boldsymbol{r_2}, \dots, \boldsymbol{r_n}, t)\, ln\big(R(\boldsymbol{r_1}, \boldsymbol{r_2}, \dots, \boldsymbol{r_n}, t)\big)$$

The argument of the logarithm term could also contain a constant factor with the units of volume to the $n^{th}$ power to compensate the units of $R$ and make the argument a dimensionless number but this constant factor will not show up in later results. The equation governing the time evolution of $R$ is a generalization of the Markov equation (125) in TSI. Equations (120−124) of TSI explain how the Markov approximation is obtained. Instead of just a single diffusion constant, $D_E$, each particle may have its own value of $D_j$ although identical particles would normally have identical values of $D$. The $n$ particle equation is

(5)

$$\partial_t R = \sum_{l=1}^{n} D_l \, \boldsymbol{\nabla}_l \cdot (\boldsymbol{\nabla}_l + \beta \boldsymbol{\nabla}_l U) R$$

where we have suppressed explicit spatial variable dependence and $\beta = 1/k_B T$. In equation (126) of TSI the proof of monotonicity used integration by parts for one variable. We will now need to treat $3n$ variables using a $3n$ dimensional generalization of the divergence theorem of $3d$ vector calculus. The proof takes the form:

(6)

$$d_t F = \int d^{3n} r \, \big(U + k_B T \, ln(R)\big) \partial_t R$$

in which a term on the right hand side integrated to zero because $R$ is normalized (and $d^{3n} r$ is shorthand for $d\boldsymbol{r_1} d\boldsymbol{r_2} \ldots d\boldsymbol{r_n}$). Substitute equation (5) into the right hand side

(7)

$$d_t F = \int d^{3n} r \, \big(U + k_B T \, ln(R)\big) \sum_{l=1}^{n} D_l \, \boldsymbol{\nabla}_l \cdot (\boldsymbol{\nabla}_l + \beta \boldsymbol{\nabla}_l U) R$$

$$= -\sum_{l=1}^{n} D_l \int d^{3n} r \, \big[\boldsymbol{\nabla}_l \big(U + k_B T \, ln(R)\big)\big] \cdot [(\boldsymbol{\nabla}_l + \beta \boldsymbol{\nabla}_l U) R]$$

$$= -\sum_{l=1}^{n} D_l \int d^{3n} r \left(\boldsymbol{\nabla}_l U + k_B T \frac{1}{R} \boldsymbol{\nabla}_l R\right) \cdot (\boldsymbol{\nabla}_l R + \beta [\boldsymbol{\nabla}_l U] R)$$

$$= -\sum_{l=1}^{n} D_l \int d^{3n} r \, \frac{k_B T}{R} (\boldsymbol{\nabla}_l R + \beta [\boldsymbol{\nabla}_l U] R) \cdot (\boldsymbol{\nabla}_l R + \beta [\boldsymbol{\nabla}_l U] R) \leq 0$$

It is clear that the decrease vanishes at full equilibrium when

(8)

$$R = \frac{1}{Q} \exp[-\beta U]$$

in which $Q$ is the normalization factor $Q = \int d^{3n} r \, \exp[-\beta U(\boldsymbol{r_1}, \boldsymbol{r_2}, \ldots, \boldsymbol{r_n})]$ for the $U$ in equation (2).

It is seen here that the Markov approximation to the contraction of the description elucidated in TSI depicts a system of $n$ coupled, interacting, diffusing particles. In the highly over-damped, very low Reynolds number, situation that this represents, non-equilibrium thermodynamics remains valid far from full equilibrium and is governed by the Helmholtz free energy which monotonically decreases with time if no outside forces or energy/matter fluxes are operating. One context in which this is an especially apt description is nano-biology as will be shown in section **III**.

## II. Least dissipation of Helmholtz free energy

In order to make the presentation as transparent as possible, we will again revert to one dimension. The Markov limit of the contracted description is given by equation (125) of TSI

(9)

$$\partial_t R(r,t) = \frac{1}{\alpha\beta}\partial_r(\partial_r + \beta U')R(r,t)$$

The coefficient in the front of the right hand side is Einstein's formula for the diffusion constant, $D$. The solution with initial condition

(10)

$$R(r,0) = \delta(r - r_0)$$

is called $P_2(r, r_0, t)$, the conditional probability density. If at time $t = 0$ the particle is at $r_0$ then the probability that it will be between $r$ and $r + dr$ at time $t$ is $dr{\times}P_2(r, r_0, t)$. This means that $P_2(r, r_0, t)$ is also the Green's function for

(12)

$$\partial_t R(r,t) = D\partial_r(\partial_r + \beta U')R(r,t)$$

where now the diffusion constant is made explicit. The formal solution to this equation is

(13)

$$R(r,t) = exp[tD\partial_r(\partial_r + \beta U')]\delta(r - r_0)$$

This isn't as useful as it appears because the two pieces of the exponentiated differential operator do not commute:

(14)

$$[D\partial_r^2, D\partial_r\beta U'] \neq 0$$

We may get around this problem using the idea of the Trotter Product formula that states

(15)

$$exp[tD\partial_r(\partial_r + \beta U')] = \lim_{n\to\infty}\left(exp\left[\frac{t}{n}D\partial_r^2\right]exp\left[\frac{t}{n}\beta D\partial_r U'\right]\right)^n$$

We will denote $\frac{t}{n}$ by $\Delta t$ and imagine that it is very small (but not as small as the intrinsic relaxation time for the non-Markov description). Now introduce the Markov property that allows us to write

(16)

$$dr_n{\times}P_2(r_n, r_0, t; path) = \prod_{j=0}^{n-1} dr_{j+1}{\times}P_2(r_{j+1}, r_j, \Delta t)$$

for $t = n\Delta t$. This means that the probability that the particle starts at $r_0$ at $t = 0$ and ends up in the interval between $r_n$ and $r_n + dr_n$ at time $t$ *having taken a particular path* is given by the left hand side, whereas the right hand side says that the particle starts at $r_0$ at $t = 0$ and at time $\Delta t$ has probability $dr_1{\times}P_2(r_1, r_0, \Delta t)$ to be in the interval between $r_1$ and $r_1 + dr_1$, and if at $r_1$ in this same spatial interval at time $\Delta t$ has probability $dr_2 \times P_2(r_2, r_1, \Delta t)$ to be in the interval between $r_2$ and $r_2 + dr_2$ at time $2\Delta t$, *et cetera*. The sequence of $r_j's$ determines the particular path, and the width of the path is determined by the sequence of $dr_j's$. We may choose the widths to be identical at each point and think of the path as a "tube" of very similar paths. This converts the probability densities into probabilities. Each factor on the right hand side of equation (16) corresponds to a factor in the Trotter Product formula

(17)

$$P_2(r_j, r_{j-1}, \Delta t) = exp\left[\Delta t D \partial_{r_j}^2\right] exp\left[\Delta t \beta D \partial_{r_j} U'\right] \delta(r_j - r_{j-1})$$

Note that the prime on the potential means the derivative with respect to $r_j$, whereas the $\partial_{r_j}$ derivative to the left of the potential term can act on the potential or past the potential. The Green's function for the left factor on the right hand side is well known and is

(18)

$$\frac{1}{\sqrt{(4\pi D \Delta t)}} exp\left[-\frac{\left(r_j - r'_{j-1}\right)^2}{4D\Delta t}\right]$$

where a primed coordinate is introduced for the purpose of computing the two factor Green's function as will become evident in a moment and is not an intermediate point in the path. Treatment of the second factor on the right hand side of equation (17) follows the macroscopic limit of the Kramers-Moyal expansion of a Master equation representation ([5] Fox, equations (I.5.10 −14)). The Green's function is

(19)

$$\delta(r'_{j-1} - \rho_{j-1}(r_{j-1}, \Delta t))$$

where $\rho_{j-1}(r_{j-1}, \Delta t)$ solves the partial differential equation (first order in time *and* first order in space)

(20)

$$\partial_t f(r'_{j-1}, t) = \beta D \partial_{r'_{j-1}} \left(U'(r'_{j-1}) f(r'_{j-1}, t)\right)$$

with initial condition $f(r'_{j-1}, 0) = \delta(r'_{j-1} - r_{j-1})$ and associated with the ordinary differential equation

(21)

$$d_t \rho_{j-1} = -\beta D U'(\rho_{j-1})$$

with initial condition $\rho_{j-1}(0) = r_{j-1}$. Multiplying equations (18) and (19) together and integrating over all $r'_{j-1}$ yields

(22)

$$P_2(r_j, r_{j-1}, \Delta t) = \frac{1}{\sqrt{(4\pi D \Delta t)}} exp\left[-\frac{\left(r_j - \rho_{j-1}(r_{j-1}, \Delta t)\right)^2}{4D\Delta t}\right]$$

Because we evolve the solution to equation (21) for a very short time, $\Delta t$, we may approximate equation (21) by expanding around $r_{j-1}$

(23)

$$d_t(\rho_{j-1} - r_{j-1}) = -\beta D\left(U'(r_{j-1}) + (\rho_{j-1} - r_{j-1})U''(r_{j-1})\right)$$

which has the solution

(24)

$$\rho_{j-1}(\Delta t) - r_{j-1} = exp[-\beta D\Delta t U''(r_{j-1})](\rho_{j-1}(0) - r_{j-1}) + \int_0^{\Delta t} ds\ exp[-\beta D(\Delta t - s)U''(r_{j-1})]\left(-\beta DU'(r_{j-1})\right)$$

$$= -\beta DU'(r_{j-1})\left(\frac{1}{\beta DU''(r_{j-1})}(1 - exp[-\beta D\Delta t U''(r_{j-1})])\right)$$

$$\cong -\beta D\Delta t U'(r_{j-1})$$

This means $\rho_{j-1}(\Delta t) = r_{j-1} - \beta D\Delta t U'(r_{j-1})$. This clearly satisfies the initial condition and exhibits the over-damped, non-ballistic motion. It permits us to rewrite equation (22)

(25)

$$P_2(r_j, r_{j-1}, \Delta t) = \frac{1}{\sqrt{(4\pi D\Delta t)}} exp\left[-\frac{\left(r_j - r_{j-1} + \beta D\Delta t U'(r_{j-1})\right)^2}{4D\Delta t}\right]$$

Equation (25), valid for sufficiently small $\Delta t$, is the central result from which all others follow.

To check that we are on the correct track, we derive here the Crooks result ([6] Crooks). Imagine a path from $r_a = r_0$ to $r_b = r_n$ along the sequence of $r_j's$. We call this the forward path. The probability for the forward path is given by

(26)

$$P^F(path) = \prod_{j=1}^{n} dr_j P_2(r_j, r_{j-1}, \Delta t)$$

in which each $dr_j$ has the same size, and the product has $j$ increasing right to left. The reverse path (here reversed in the sequence of coordinates only since there are no explicit momenta) has probability

(27)

$$P^R(reverse\ path) = \prod_{j=1}^{n} dr_{j-1} P_2(r_{j-1}, r_j, \Delta t)$$

in which the product has $j$ decreasing right to left, and

(28)

$$P_2(r_{j-1}, r_j, \Delta t) = \frac{1}{\sqrt{(4\pi D\Delta t)}} exp\left[-\frac{\left(r_{j-1} - r_j + \beta D\Delta t U'(r_j)\right)^2}{4D\Delta t}\right]$$

Especially note the different points at which the forces are evaluated for forward and reverse paths as follows directly from the derivation for the reverse path that parallels the forward path derivation. Now for Crooks' seminal observation

(29)

$$ln\left(\frac{P^F(path)}{P^R(reverse\ path)}\right) = \sum_{j=1}^{n} ln\left(\frac{P_2\left(r_j - r_{j-1} + \beta D\Delta t U'(r_{j-1})\right)}{P_2\left(r_{j-1} - r_j + \beta D\Delta t U'(r_j)\right)}\right) =$$

$$\sum_{j=1}^{n}\left(-\frac{1}{4D\Delta t}\left(\left(r_j - r_{j-1} + \beta D\Delta t U'(r_{j-1})\right)^2 - \left(r_{j-1} - r_j + \beta D\Delta t U'(r_j)\right)^2\right)\right) =$$

$$\sum_{j=1}^{n}\left(-\frac{1}{2}\beta(r_j - r_{j-1})\left(U'(r_{j-1}) + U'(r_j)\right)\right)$$

where all the $dr_j's$ have cancelled out (except for $dr_0$ and $dr_n$ which were assumed equal just below equation (26)) and terms of order $\Delta t^2$ have been dropped. For paths of closely spaced points we may write

(30)

$$U'(r_j) = U'\left(\frac{1}{2}(r_j + r_{j-1}) + \frac{1}{2}(r_j - r_{j-1})\right) =$$

$$U'\left(\frac{1}{2}(r_j + r_{j-1})\right) + \frac{1}{2}(r_j - r_{j-1})U''\left(\frac{1}{2}(r_j + r_{j-1})\right)$$

and

$$U'(r_{j-1}) = U'\left(\frac{1}{2}(r_j + r_{j-1}) - \frac{1}{2}(r_j - r_{j-1})\right) =$$

$$U'\left(\frac{1}{2}(r_j + r_{j-1})\right) - \frac{1}{2}(r_j - r_{j-1})U''\left(\frac{1}{2}(r_j + r_{j-1})\right)$$

Neglecting coordinate differences of second order or higher we have, in the large $n$ limit

(31)

$$ln\left(\frac{P^F(path)}{P^R(reverse\ path)}\right) = -\beta\sum_{j=1}^{n}(r_j - r_{j-1})U'\left(\frac{1}{2}(r_j + r_{j-1})\right) \to -\beta(U(b) - U(a))$$

This change in internal potential energy (an obligatory decrease for over-damped motion) converts to a heat input into the reservoir (water). $\Delta Q_{res} \equiv -\Delta U$ and implies Crooks' identity

(32)

$$\frac{P^F(path)}{P^R(reverse\ path)} = exp[\beta\Delta Q_{res}]$$

Note that the integration of the work done in equation (31) involves the midpoint rule, consistent with the Stratonovich interpretation of stochastic differential equations with white noise, such as underlies the incorporation of water molecule effects as Brownian motion for the solute molecules (see TSI). Reduction of a non-Markov picture to a Markov picture always leads to the Stratonovich interpretation in a natural way as has occurred here ([5] Fox).

Some authors have interpreted the left hand side of equation (32) as a measure of a path's irreversibility, in the case of full phase space trajectories. They point out that a positive $\Delta Q_{res}$ means the forward path is more probable than the reverse path. Then they say that for $\Delta Q_{res} = 0$ the path is reversible and, therefore, the process is reversible. A look at pure diffusion will demonstrate that this interpretation is misleading. Diffusion is the quintessential irreversible process in physics. However, when we set $U \equiv 0$ in equation (29) there is a perfect cancellation and the ratio of forward path probability to reverse path probability is one. This doesn't mean the process of diffusion is reversible. It only means that the path entropy in the reverse direction matches that for the forward direction. The forward path entropy is

(33)

$$S^F(path) = -k_B \sum_{j=1}^{n} \left( \ln\left( \frac{dr_j}{\sqrt{(4\pi D\Delta t)}} \right) - \frac{\left(r_j - r_{j-1}\right)^2}{4D\Delta t} \right) \geq 0$$

which is positive, since the coordinate differentials are smaller than the standard deviations, and reflects the irreversible property for diffusion. The expression for the reverse path is the same (remember $dr_0$ and $dr_n$ are the same size). Thus going from $r_a$ to $r_b$ increases the path entropy and so does going from $r_b$ to $r_a$, increasing the path entropy by the same amount. If we did not restrict the tube of paths over its entire range, the spread in positions for the forward paths would be much greater at the end than at the beginning, and starting from the points of the final distribution would not shrink the distribution down to the initial distribution for the forward paths, as the final distribution for the reverse paths, but would instead spread it out even more. For diffusion the forward and reverse bundles of restricted paths have the same probabilities, and their path entropies are both positive as befits an irreversible process.

Equation (25) suggests another immediate result. Consider the most probable path between two points $r_a$ and $r_b$. This means we want the path for which

(34)

$$P^F(path) = \prod_{j=1}^{n} dr_j P_2\left(r_j, r_{j-1}, \Delta t\right) \; is\ a\ maximum$$

where $r_0 = r_a$ and $r_n = r_b$. Since all probabilities are positive, the maximum occurs when the logarithm is a maximum. Using equation (25) this means

(35)

$$\max \ln\left(P^F(path)\right) = max \sum_{j=1}^{n}\left(\ln\left(\frac{dr_j}{\sqrt{(4\pi D\Delta t)}}\right) - \frac{\left(r_j - r_{j-1} + \beta D\Delta t U'(r_{j-1})\right)^2}{4D\Delta t}\right)$$

Thus the most probable path is determined by the condition (all $dr_j's$ are the same)

(36)

$$\frac{1}{4D\Delta t}\sum_{j=1}^{n}\left(r_j - r_{j-1} + \beta D\Delta t U'(r_{j-1})\right)^2 \ is\ a\ minimum$$

subject to the constraint

(37)

$$r_n - r_0 = r_b - r_a$$

For sufficiently small $\Delta t$ the coordinate term dominates the potential term and the minimum can be found, using a Lagrange multiplier, $\lambda$, for the constraint, and is given by

(38)

$$r_j - r_{j-1} = -\frac{\lambda}{2}$$

where

$$-\frac{n}{2}\lambda = r_b - r_a$$

This is exactly the result for pure diffusion without inter-particle potentials. As will be argued in the next section, typical values for the physical parameters appearing above for intermediate sized solute molecules (think ubiquinone) inside a cell at room temperature are

(39)

$$D \cong 10^{-7}\frac{cm^2}{s}$$

$$\beta \cong \frac{1}{4}\times 10^{14}\frac{s^2}{g - cm^2}$$

$$\Delta t \cong 10^{-12}s$$

$$U' \cong pN = 10^{-7}\frac{g - cm}{s^2}$$

The product of these factors is $2.5 \times 10^{-12}\ cm$, whereas the $r_j$ spacing may well be angstroms. If we include terms from the potentials, we get a more complicated set of equations for the minimum with constraint. After a bit of algebra we get (dropping terms of second order in $\Delta t$ and/or $r_j - r_{j-1}$)

(40)

$$0 = \sum_{j=1}^{n}(r_{j-1} - r_j)\beta D\Delta t U''(r_{j-1})$$

$$r_1 - r_0 + \beta D\Delta t U'(r_0) = -\frac{\lambda}{2}$$

$$r_j - r_{j-1} + \beta D \Delta t U'(r_{j-1}) = r_{j-1} - r_{j-2} + \beta D \Delta t U'(r_{j-2})$$
$$= -\frac{\lambda}{2} \; for\, j \in [2, n]$$
$$r_n - r_{n-1} + \beta D \Delta t U'(r_{n-1}) = -\frac{\lambda}{2}$$
$$-\frac{n}{2}\lambda = r_b - r_a + \sum_{j=0}^{n-1} \beta D \Delta t U'(r_j)$$

These equations require a numerical solution in general. When the potentials are absent they reduce to the pure diffusion equations (38) above. Because the potential dependent terms are small according to equation (39) the shifts in the locations of the $r_j\,'s$ for the most probable path from $r_a$ to $r_b$ are also small. Any other forward path from $r_a$ to $r_b$ that deviates from the path of points that minimizes the expression in equation (36) will decrease the probability for this other forward path. This amounts to increasing the system entropy production relative to the most probable path (the system entropy production for the path is given by multiplying the logarithm of the path probability given in equation (35) by minus Boltzmann's constant), and since the potential energy change between $r_a$ and $r_b$ remains unchanged for conservative forces (i.e. is path independent, in contrast to the path entropy that is irreversible for interactive diffusing particles and, therefore, path dependent), the amount of Helmholtz free energy dissipated is increased. Put another way, *the Helmholtz free energy dissipation for the most probable path between two points is least.* Conversely, *the path between two points that dissipates the Helmholtz free energy the least is the most probable path.*

Three remarks are in order. 1) In the generalization to $n$ particles, each in 3 dimensions, the sum of squares as in equation (36) is preserved. The idea of conservative forces is also maintained in $3n$ dimensions. Thus the minimization with constraint follows directly. 2) If one explicitly includes the effects of the potential terms on the minimization the result is that the spacing of the coordinate differences is no longer uniform as in equation (38) but contains variable, order $\Delta t$ corrections given by equation (40). Nevertheless the sum of the squares is a minimum and all other paths will generate more system entropy than the most probable path that dissipates the Helmholtz free energy the least. For related results see ([7] English) and references therein. 3) Ubiquinone has a radius of $0.75\ nm$, and the standard deviation for the Gaussian in equation (25) is $\sqrt{2D\Delta t}$. For the values given in equation (39) this works out to be $4.5 \times 10^{-10}\ cm$ an amount smaller than the radius. Choosing a larger $\Delta t$, say $10^{-8}\ s$, makes the standard deviation $0.45\ nm$. A single step in a tube of paths within one standard deviation of the most probable path accounts for 68% of the probability for that step.

We have treated the case of the most probable path from $r_a$ to $r_b$. We may also ask *what is the most probable path starting from* $r_a$, where the end point $r_b$ is determined by the path itself. The result of this question is given by equation (36). One simply sets each quadratic term equal to zero. For pure diffusion this means that, most probably, the particle stays put at its initial position. With inter-particle potentials present, the system simply follows the deterministic over-damped trajectory determined by first order ordinary differential equations in $3n$ dimensions, like equation (21) in one dimension. Thus the most probable path, given only the starting point, is the deterministic over-damped path. All other paths starting from $r_a$ generate more path entropy than the deterministic, over-damped path and may end up at different final points. The most probable path between two specified points is something different and is the path with the least dissipation of Helmholtz free energy compared to all other paths between the same two points. *In both cases the most probable path dissipates the least Helmholtz free energy, and conversely.*

## III. Implications for nano-biology

*Nano-biology* takes place on a length scale of nanometers, or 10 angstroms. This is the same as for *molecular biology* and for *biochemistry*. It is a very different realm compared to the macroscopic world. At the nano-scale viscosity dominates everything dynamical and inertia has no significant effects. As shown by Howard Berg ([3] Berg, [2] Fox) an *E. Coli* bacteria propelled by a flagella can achieve speeds of $2 \times 10^{-3} cm/s$ for a run of about 1 $s$. The bacteria's length is 2 microns (2,000 nanometers), its width is one micron and the viscosity of its natural environment is 0.027poise. The Reynolds number is $7 \times 10^{-6}$. These are the parameters of a regime in which viscosity dominates inertia. If the bacterium is swimming at full speed and the flagella motor is suddenly turned off the bacterium comes to a stop in just $10^{-10}$ $cm$! In contrast to this situation for *E. Coli* an 8 centimeters long minnow swimming in a pond has a Reynolds number around 20,000. Viscosity is a minor drag for the minnow. The differences between an *E. Coli* and a minnow are very great. These two entities are on the mesoscopic (microns, $\mu$) and macroscopic (centimeters, cm) size scales respectively. On the truly microscopic (nanometers, nm) scale is the molecule ubiquinone.

Ubiquinone is a non-protein catalyst centrally located in the electron transport chains of mitochondria and chloroplasts (plastoquinone) ([2] Fox, chapter 2). It connects the iron-sulfur protein segment of the electron transport chain to the cytochrome (heme-iron) protein segment of the chain, as well as transferring protons from inside the membrane to outside the membrane. It becomes reduced (2 electrons and 2 protons) on the inside of the membrane and oxidized (2 electrons and 2 protons) on the outside thereby coupling the redox energy of energy metabolism to the proton energy of the membrane (chemiosmosis). Ubiquinone has a molecular weight of 862 (864) daltons and a radius of 0.75 nm. It moves by Brownian motion inside the lipid membrane that has a viscosity of 25 cp (centipoise, or 0.01 g/cm-s). By randomly changing sides, and either reducing or oxidizing, depending on which side it is on, it on average engages in a catalytic redox cycle, vectorially coupling intra-membrane electron current to trans-membrane proton current. This motion is strongly over-damped with a Reynolds number of $2.4\times10^{-6}$. The Langevin relaxation time is $4\times10^{-15}s$. Typically one diffusive transit of the membrane width by ubiquinone takes on average $10^{-6}s$, a time very much longer than the relaxation time and therefore in the extreme Markov limit of this naturally non-Markov contracted process. Treating the ubiquinone dynamics as a pure diffusion process is justified by this strong separation of time scales ([2] Fox). This circumstance will hold for the myriads of smaller molecules inside a cell, including amino acids, sugar molecules, nucleosides and ATP. The dimensions of a typical cell are small enough for diffusion to be a robust source of motion that can transfer any small molecule across the entire cell in sub-microsecond times.

The motion of kinesin, a motor protein, on microtubules has also been analyzed by these methods ([2] Fox, chapter 4). The moving “heads” of kinesin use Brownian motion for their movement. In this case the kinesin heads are attached to the kinesin “tether” by “neck linkers” that exhibit elasticity. Inclusion of elasticity in the analysis introduces another relaxation time scale as well as an elasticity period but it works out that for all three time scales the dynamics is in the strongly over-damped Markov limit (contrary to what a major kinesin experimentalist has claimed).

The preceding remarks are aimed at justification for the Markov limit of the non-Markov contracted description. In most cases it works out that the system is strongly over-damped, has very small Reynolds number and the dynamics is dominated by viscosity. Nevertheless, on the length scales involved, Brownian motion, or its strongly over-damped equivalent, diffusion, is very robust as a source of motion. At this stage let us look at the evolutionary significance of the least dissipation of Helmholtz free energy principle. Consider the basic core process of energy metabolism, the oxidation of sugar by oxygen. Chemically, there are many sugar sizes and isotopes for given size. The basic formula for this circumstance is $(H_2CO)_n$. For example the sugar glucose that is used in glycolysis has $n = 6$. There are many stereo-isotopes with $n = 6$. We will omit further concern for this variety. Oxidation of glucose can be represented by the formula
(41)

$$(H_2CO)_6 + 6\,O_2 \leftrightarrow 6\,CO_2 + 6\,H_2O + energy$$

If the glucose is burned in a clean flame the conversion is total and the energy released is pure heat amounting to a Gibbs free energy change of $-686.9\ kcal/mol$. The free energy of formation of carbon dioxide and of water is much more negative than that for an equivalent amount of glucose (the oxygen free energy of formation is zero). The Boltzmann-Planck formula for the amount of glucose left in equilibrium, at standard temperature and pressure, is so unfavorable that for a macroscopic amount of glucose we could handle and manipulate there would not be left even one molecule of glucose.

(42)

$$exp\left[-\frac{687{,}000}{594}\right] = exp[-1157] = 3.3\times 10^{-503}$$

The denominator is $RT$ in units of cal/mol and at 300 Kelvin. Let the initial state be one with equivalent amounts of glucose and oxygen molecules. This is a non-equilibrium state. Let the final state be the state in which all glucose has been oxidized into carbon dioxide and water. If nothing else is present then the most probable path will be in essence the only type of path possible, an oxidation of glucose gone to completion. Conversely, the spontaneous formation of glucose and molecular oxygen from carbon dioxide and water is extremely unlikely.

Over the course of evolution energy metabolism has become more and more complex and its ability to harness useful free energy from the oxidation of glucose has increased commensurately. A very major need for energy is caused by the synthesis of the macromolecules, proteins, polynucleotides and polysaccharides. In the case of the first two types, the reason is that as monomers are added to the growing polymer chain dehydration linkages are made. Because of the abundance of water this direction of polymer synthesis is strongly thermodynamically inhibited. Cells need to be able activate the monomers in order to override the strong hydrolysis tendency. The universal mechanism to do this is some variation on phosphorylation of the monomers, called activation, and the ultimate source for phosphate potential is ATP. (Proteins are made from amino acids that are activated into amino-acyl-adenylates by ATP. Polynucleotides are made from nucleosides (monophosphates) that are activated into nucleotides (triphosphates) by ATP.) The core of energy metabolism, the slow oxidation of glucose by oxygen, produces the needed ATP ([8] Fox). Four stages of glucose oxidation have evolved. The most primitive is glycolysis that converts one glucose into two pyruvates, two ATP's and the reducing equivalent of two NADH's. The two ATP's only account for slightly more than 2% of the available free energy. Next comes the conversion of pyruvate into acetyl-CoA. This stage releases two $CO_2$ and two NADH's, as well as generating the versatile thioester acetyl-CoA. The third stage is the citric acid cycle that produces large amounts of reducing potential in the forms of NADH and $FADH_2$. These reducing equivalents release electrons into the electron transport chains, in the fourth stage, where coupling of electron currents to proton currents results in a transmembrane electrochemical potential (chemiosmosis). This proton potential is discharged by ATPase complexes embedded in the membrane that convert proton current into ATP synthesis. When all stages operate, such as in ourselves, in our mitochondria, 38% of the energy available from glucose oxidation is retained, for the biosynthesis of polymers, in the form of ATP (actually 34 ATP's and 2 GTP's). The question is: why did evolution take glucose oxidation in this direction, so that a greater amount of useful energy is harvested, instead of simply converting all those glucose kcal/mol directly into heat?

The answer to the question that ended the preceding paragraph is that the most probable path is one that minimizes the decrease in Helmholtz free energy (We spoke of Gibbs free energy in the preceding paragraph. The difference from Helmholtz free energy involves a pressure-volume term that remains largely invariant in the background for our aqueous cells. Alternatively we could generalize our least dissipation of free energy principle to the Gibbs case with the pressure-volume term playing a role more akin to the internal energy than to the entropy.). If other molecular species are present, such as catalysts, then more chemical options can exist and the most probable paths will involve less free energy dissipation and more production of intermediates that retain some free energy. Catalysts are regenerated by catalytic cycles. In the purely classical modeling provided by the coupled interacting and diffusing particles described by equation (5) catalysts can be modeled by including three body potentials. Interactions between species A and B can be described by a two body potential. If their interaction is to be affected by a catalyst, a three body potential can perform that function by changing the two body, A/B interaction, when the catalyst is near. For example, A and B may have a short ranged weak attraction, with a deep localized well separated from the attraction by a barrier. In over-damped dynamics A and B would not likely get over the barrier and into the well (a deep well could model bonding). The purely deterministic over-damped motion is absolutely prevented from going over the barrier. Catalyst C, when close enough to A and B, can affect the effective two body interaction between A and B by reducing the barrier and catalyzing access to the well, through an appropriately constructed three body potential. If desired even the approximation of the particles as points could be relaxed so that the particles have volume, asymmetry and angle coordinates for orientation. All of this makes the description very complex but remains possible in principle. With these preliminaries let us consider the evolution of a primitive energy metabolism pathway that invokes the modern biological state.

One primitive source of reducing electrons is the UV excitation of ferrous iron, plentiful on the early earth. This is a most basic redox process:
(43)

$$UV + Fe^{2+} \rightarrow Fe^{3+} + e^{-}$$

Thiols are molecules containing a reduced sulfur group bonded to a carbon atom
(44)

$$R - SH$$

in which $R$ denotes the rest of the molecule. Glyceraldehyde is a triose sugar $(H_2CO)_3$
(45)

$$\begin{array}{ccccccc} & & & & & & \mathrm{O} \\ & \mathrm{H} & & \mathrm{H} & & & \| \\ \mathrm{H} & \mathrm{C} & - & \mathrm{C} & - & & \mathrm{C} \\ & \mathrm{O} & & \mathrm{O} & & & \mathrm{H} \\ & \mathrm{H} & & \mathrm{H} & & & \end{array}$$

Using the oxidizing power of ferric iron, a thioester can be formed with glyceraldehyde

(46)

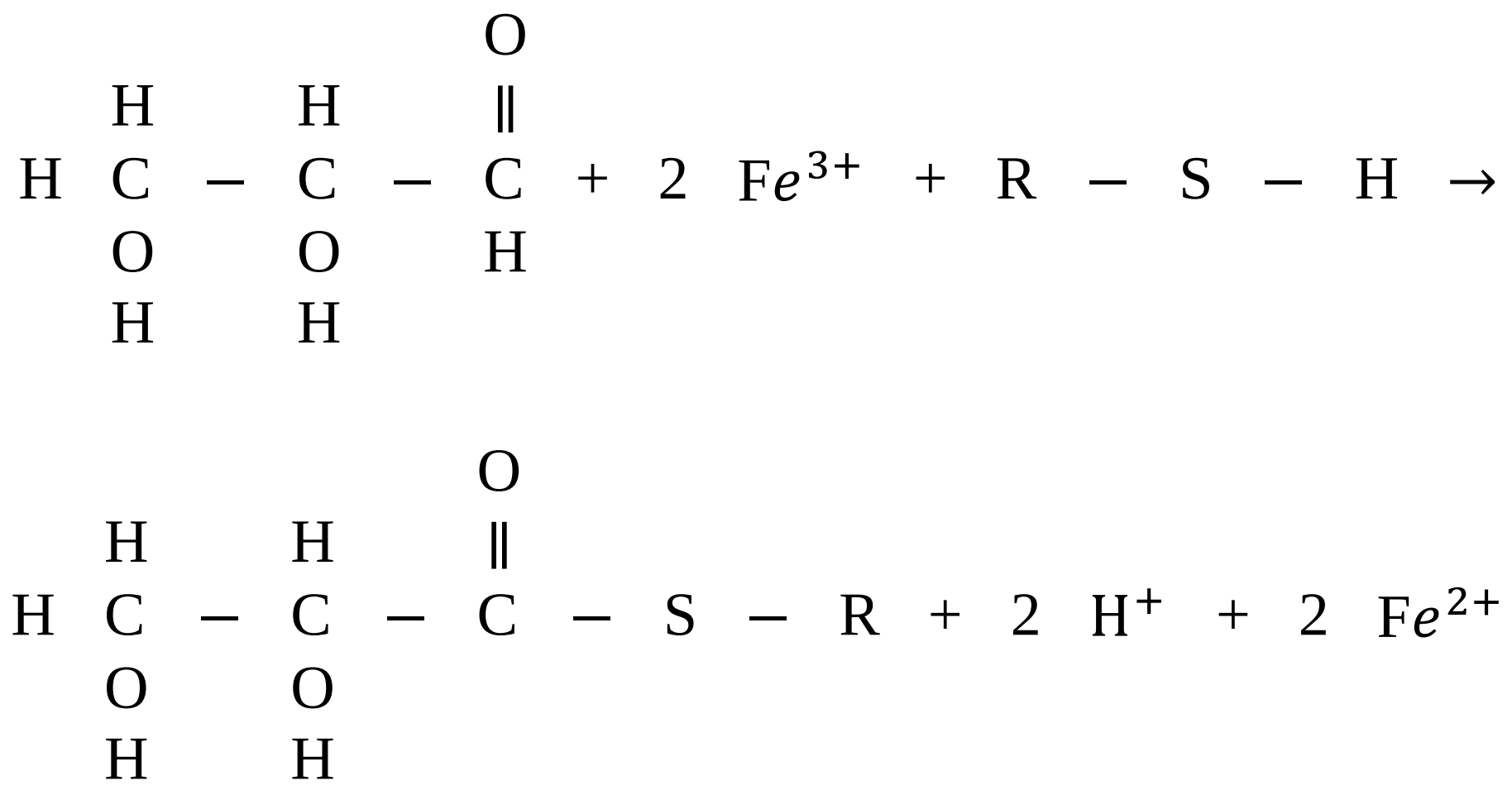

This is an energy rich molecule, retaining free energy derived from the UV oxidation of iron. This thioester can react with inorganic phosphate to form energy rich acylphosphate.

(47)

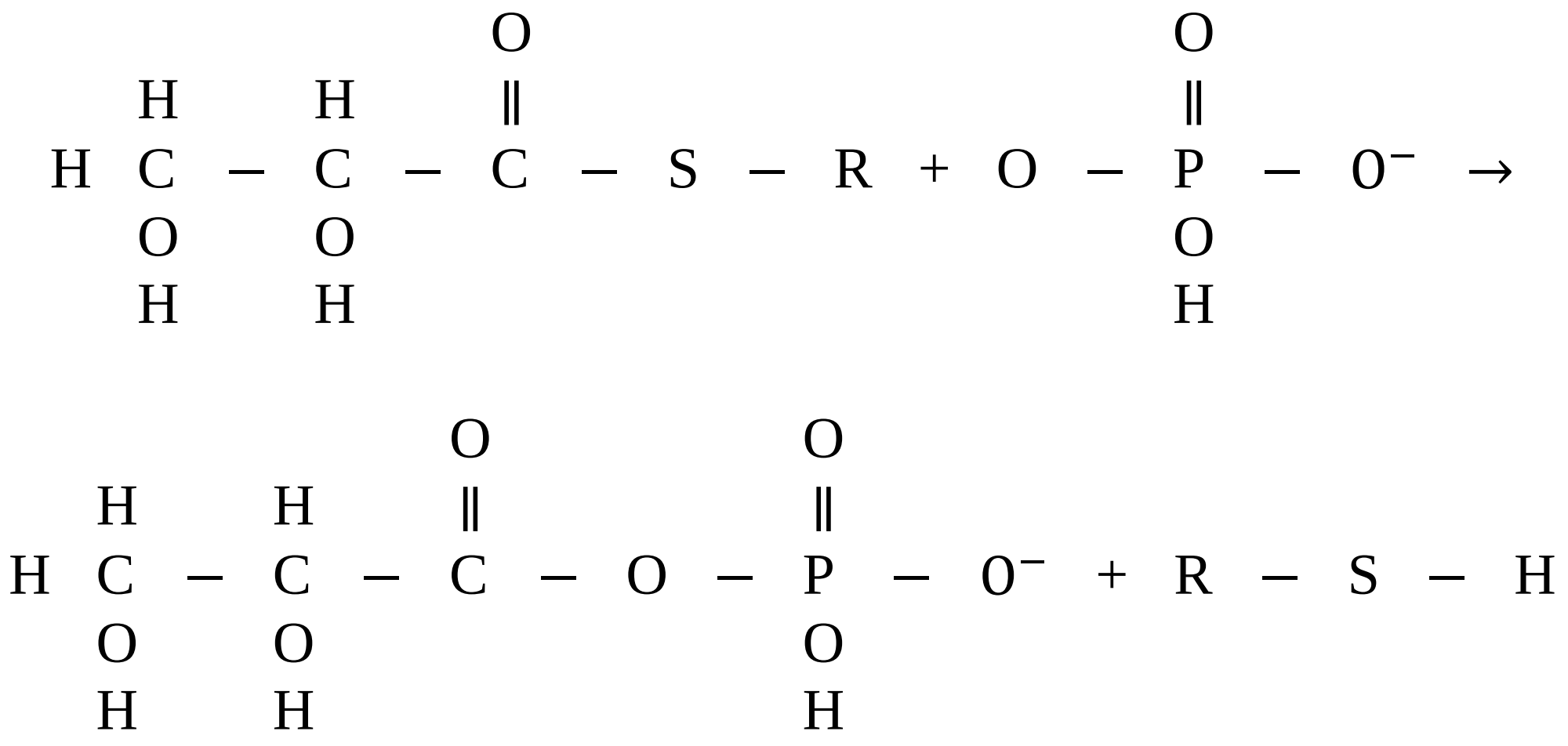

The thiol has been regenerated like a catalyst. Indeed, in glycolysis there is an enzyme step involving the action of a thiol group in its active site that these primitive steps here imitate. This imitation can be taken a step further by letting the acylphosphate react with another molecule of inorganic phosphate to form the energy rich pyrophosphate that is capable of activating monomers.

(48)

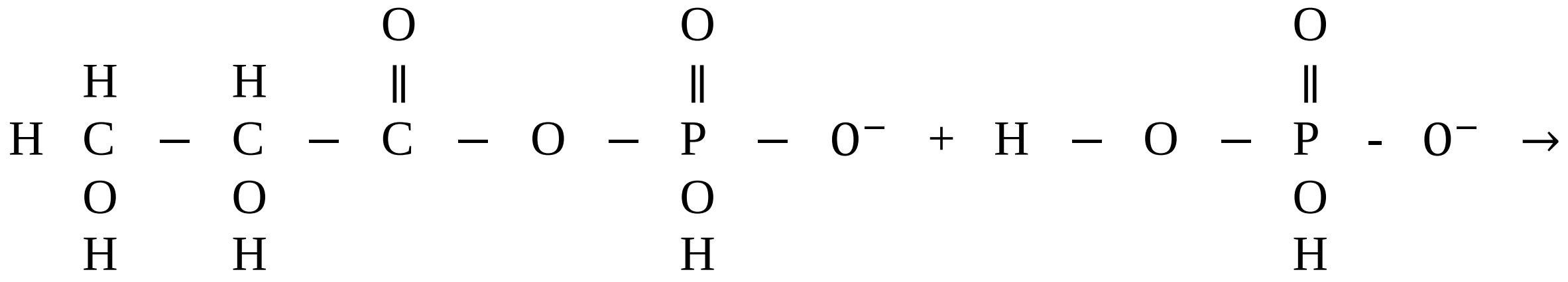

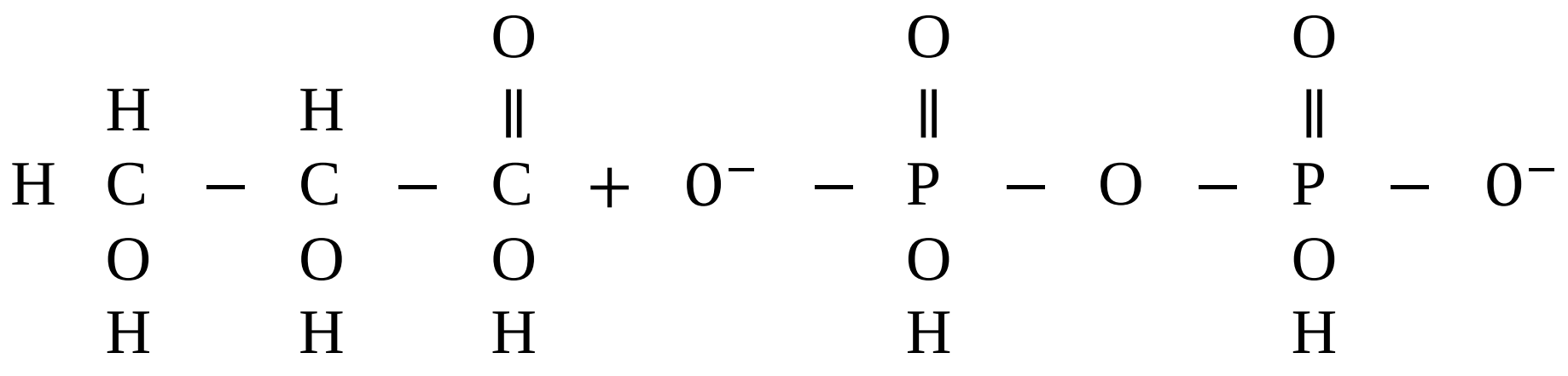

The initial glyceraldehyde ends up partially oxidized as glycerate. These steps have been demonstrated in the laboratory without specific catalysts ([9] Weber, [10] de Duve). With pyrophosphate present activation and polymerization of amino acids becomes possible, and is more probable than the direct hydrolysis of the pyrophosphate. Similar considerations regarding the replication and transcription of DNA follow *mutatis mutandis*.

Thioesters, such as in equation (46), are subject to hydrolysis. Formation of an acylphosphate instead of direct hydrolysis preserves some of the original free energy. By our principle of least dissipation formation of the acylphosphate is more probable than hydrolysis of the thioester. Similarly, the acylphosphate is subject to hydrolysis but instead can from pyrophosphate, an energy rich precursor to monomer activation. Once again the formation of pyrophosphate is more probable because it preserves free energy. We are beginning to see how and why many steps have arisen in the modern metabolic harvesting of energy from the oxidation of glucose. The most probable events are those that preserve free energy thereby dissipating it the least. As possibilities to preserve free energy arise they are most probable. This circumstance has been established here for strongly over-damped, thermostated systems, such as living cells. It has been justified for the Markov limit of a purely spatial contracted description using projection operators, boson operator representations of differential operators, the Trotter product formula and Green's functions.

## References

[1] R. F. Fox, arXiv:1409.5712, http://www.fefox.com/ARTICLES/CTTS.pdf
[2] R. F. Fox, Nanotech 2004, **3**, 18 (2004); Rectified Brownian Motion, an eBook, http://www.fefox.com/ARTICLES/RectifiedBrownianMotioneBook.pdf
[3] H. Berg, *Random Walks in Biology, expanded edition*, (Princeton U. Press, Princeton, N. J. 1993)
[4] U. R. Steiger and R. F. Fox, *Journal of Mathematical Physics* **23**, 296-314 (1982).
[5] R. F. Fox, *Physics Reports C* **48**, 179-283 (1978).
[6] G. E. Crooks, *Physical Review E* **60**, 2721 (1999).
[7] J. L. English, *The Journal of Chemical Physics* **139**, 121923 (2013).
[8] R. F. Fox, http://www.fefox.com/ARTICLES/Energy_metabolism_overview.pdf
[9] A. L. Weber, *Origin of Life, Evolution of the Biosphere*, 17, 107 (1987).

[10] C. de Duve, *Blueprint for a Cell: The Nature and Origin of Life*, (Neil Paterson Pubs., Burlington, North Carolina, 1991).

March 14, 2015
Ronald F. Fox
Smyrna, Georgia